\documentclass{PoS}

\PoS{PoS(LAT2005)032}

\title{The hadron spectrum from twisted mass QCD with a strange quark }

\ShortTitle{The hadron spectrum from twisted mass QCD with a strange quark }

\author{Abdou M. Abdel-Rehim$^a$, \speaker{Randy Lewis}$^a$
        and R. M. Woloshyn$^{b}$ \\ \\
        \llap{$^a$}Department of Physics, University of Regina,
        Regina, SK, Canada, S4S 0A2\\
        \llap{$^b$}TRIUMF, 4004 Wesbrook Mall, Vancouver, BC, Canada, V6T 2A3
        \\ \\
        E-mail: \email{randy.lewis@uregina.ca}
        }

\abstract{
Various suggestions exist for incorporating the strange quark into twisted
mass QCD.  One option for quenched simulations is to employ two twisted
doublets, $(u,d)$ and $(c,s)$, with separate twist angles.  Working in the
isospin limit, preliminary results for the quenched spectrum of strange
hadrons are presented, with both twist angles tuned to $\pi/2$.  Splittings
within the mass multiplets provide some insight into the symmetry breaking
effects of twisted mass lattice QCD.}

\FullConference{XXIIIrd International Symposium on Lattice Field Theory\\
		 25-30 July 2005\\
		 Trinity College, Dublin, Ireland}

\begin{document}

\section{Motivation}

Twisted mass lattice QCD (tmLQCD) offers an efficient mechanism for
eliminating unphysical zero modes\cite{Frezzotti:2000nk} and removing $O(a)$
errors\cite{Frezzotti:2003ni} from Wilson simulations.
Given the successes of tmLQCD in dealing with the $(u,d)$ quark doublet,
it now remains to extend the theory to include other quark flavours,
noting that the next lightest pair, $(c,s)$, is far from being degenerate.
The present work makes a step in this direction by investigating one option.

Some unphysical flavour symmetry violation occurs at non-zero lattice spacings
in tmLQCD.  For example, in the two flavour theory with a degenerate $(u,d)$
doublet,\cite{Abdel-Rehim:2005gz,Walker-Loud:2005bt}
\begin{equation}
\left(m_{\Delta^{++}}=m_{\Delta^-}\right)
 ~~~\neq~~~ \left(m_{\Delta^0}=m_{\Delta^+}\right).
\end{equation}
When tmLQCD is extended to include the strange quark,
many splittings arise among strange hadrons as well.
Since all unphysical flavour splittings must vanish in the continuum limit,
they provide a measure of discretization errors in tmLQCD.

\section{Lattice action}

There is more than one way to incorporate the strange quark into tmLQCD.
In the present work, we use a tmLQCD fermion action similar to that of
Ref.~\cite{Pena:2004gb} which is a four-quark action, though the fourth
(charm) quark will not affect any of our quenched three-flavour
spectrum computations.  The action is
\begin{equation}
S_f[\psi,\bar\psi,U] = a^4\sum_x\bar\psi(x)\left(M+i\mu\gamma_5
      +\gamma\cdot\nabla^\pm
      -\frac{a}{2}\sum_\nu\nabla^*_\nu\nabla_\nu\right)\psi(x),
\end{equation}
where
\begin{eqnarray}
\psi^T &=& (u,d,c,s), \\
M &=& diag(M_l,M_l,M_c,M_s), \\
\mu &=& diag(\mu_l,-\mu_l,\mu_c,-\mu_s).
\end{eqnarray}
The parameters $M_n$ and $\mu_n$ denote the standard quark masses and twisted
quark masses respectively.
Forward ($\nabla$), backward ($\nabla^*$) and symmetric ($\nabla^\pm$)
derivatives are standard.
To obtain automatic $O(a)$ improvement we choose maximal twist.
The classical definition would be
\begin{eqnarray}
\omega_1 &\equiv& \arctan\left(\frac{\mu_l}{M_l}\right) = \frac{\pi}{2}, \\
\omega_2 &\equiv& \arctan\left(\frac{\mu_c}{M_c}\right)
                = \arctan\left(\frac{\mu_s}{M_s}\right) = \frac{\pi}{2},
\end{eqnarray}
but quantum corrections lead to renormalization of both $M$ and $\mu$.
Maximal twist is therefore defined nonperturbatively
by\cite{Abdel-Rehim:2005gz,Farchioni:2004fs,Sharpe:2004ny}
\begin{equation}\label{maximaltwist}
\tan\omega_n = \left(
               \frac{i\sum_{\vec x}\left<\tilde V_\nu^-(\vec x,t)P^+(0)\right>}
                     {\sum_{\vec x}\left<\tilde A_\nu^-(\vec x,t)P^+(0)\right>}
               \right)_n
\end{equation}
where a tilde denotes currents constructed from fields in the twisted basis.
The correlators on the right hand side of Eq.~(\ref{maximaltwist})
require only a single fermion propagator, hence the single subscript $n$.
In the present work, this fermion action will be combined with the
standard Wilson gauge action.

\section{Simulation details}

\begin{table}
\begin{center}
\begin{tabular}{cccccl}
\hline
$\beta$ & \#sites & \#configs & $aM_q$ & $a\mu_q$ & physical mass \\
\hline
5.85 & $20^3\times40$ & 594 & -0.8965 & 0.0376 & $\sim m_s$ \\
     &                &     & -0.9071 & 0.0188 & $\sim m_s/2$ \\
     &                &     & -0.9110 & 0.01252 & $\sim m_s/3$ \\
     &                &     & -0.9150 & 0.00627 & $\sim m_s/6$ \\
6.0 & $20^3\times48$ & 600 & -0.8110 & 0.030 & $\sim m_s$ \\
    &                &     & -0.8170 & 0.015 & $\sim m_s/2$ \\
    &                &     & -0.8195 & 0.010 & $\sim m_s/3$ \\
    &                &     & -0.8210 & 0.005 & $\sim m_s/6$ \\
\hline
\end{tabular}
\end{center}
\vspace{-1mm}

\caption{Parameter choices for the simulations.}\label{latticedetails}
\end{table}

Table~\ref{latticedetails} shows the parameter values chosen for our action
and lattices.
The authors of Ref.~\cite{Jansen:2003ir} used $r_0$ to determine the lattice
spacings: $a=0.123$ fm for $\beta=5.85$ and $a=0.093$ fm for $\beta=6.0$.

The standard mass parameter values, $aM_q$, were determined in
Ref.~\cite{Abdel-Rehim:2005gz} to correspond to maximal twist as defined by
Eq.~(\ref{maximaltwist}).
The twisted mass parameters range from the approximate physical strange quark
mass, $m_s$, to 1/6 of $m_s$.
Notice that the standard mass parameter is tuned for each individual
twisted mass parameter so,
for example, the quark and anti-quark in a computation of the kaon mass will
have separate $aM_q$ values.

Local interpolating operators are used throughout this work, and
three state fits to the resulting correlators are performed.

\section{Results for meson masses}

\begin{figure}
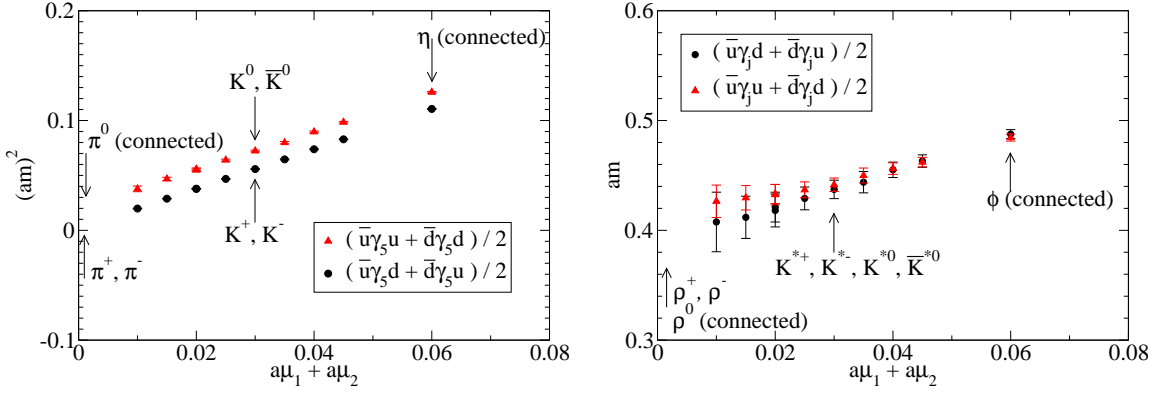

\hspace{-4mm}
\scalebox{0.29}{\includegraphics*[-5mm,11mm][257mm,19cm]{meson1b.eps}}
\scalebox{0.29}{\includegraphics*[-5mm,11mm][257mm,19cm]{meson2b.eps}}
\caption{Pseudoscalar (left panel) and vector (right panel) meson masses
         as functions of the sum of quark and anti-quark twisted
         masses at $\beta=6.0$.}\label{mesonplots}
\end{figure}

Numerical results for ground state pseudoscalar meson and vector meson masses
are displayed for $\beta=6.0$ in Fig.~\ref{mesonplots}.
(Comparable results were obtained for $\beta=5.85$.)  All quark/anti-quark mass
combinations from Table~\ref{latticedetails} are shown in this figure, with
arrows indicating the values of twisted quark masses
(i.e. the horizontal axis of each plot) corresponding to the
physical hadrons.
For pseudoscalar mesons, significant mass splittings are evident between
charged and neutral mesons, as read from the vertical axis of each plot.

\begin{figure}
\begin{center}
\scalebox{0.4}{\includegraphics*[0cm,11mm][26cm,19cm]{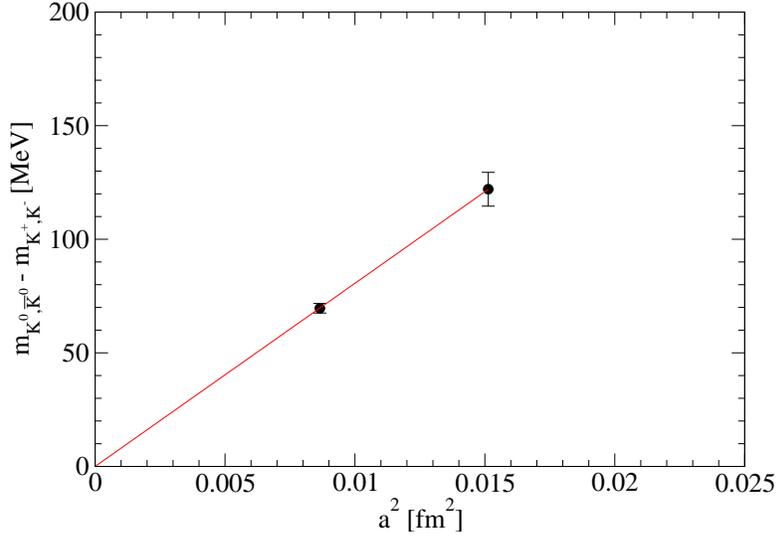}}
\caption{Mass splitting between neutral and charged kaons as a function of
         squared lattice spacing.}\label{kaonplot}
\end{center}
\end{figure}

The observed mass splitting between $\pi^\pm$ and $\pi^0$ is affected both by
quenching and by our omission of ``disconnected'' contributions.
(See Ref.~\cite{Farchioni:2005hf} for a recent tmLQCD study of this splitting.)
However, the $K^\pm$ to $K^0,\bar K^0$ splitting requires no disconnected
contributions and thus our data provide the complete quenched result in that
case.  Given the automatic $O(a)$ improvement of masses in
tmLQCD\cite{Frezzotti:2003ni}, this unphysical kaon splitting should vanish
linearly with $a^2$.  Figure~\ref{kaonplot} shows that our two $\beta$
values are consistent with this expectation.

In contrast to the pseudoscalar mesons, Fig.~\ref{mesonplots} shows no
significant splitting between charged and neutral $K^*$ mesons within
the statistical uncertainties.

\section{Results for baryon masses}

\begin{figure}
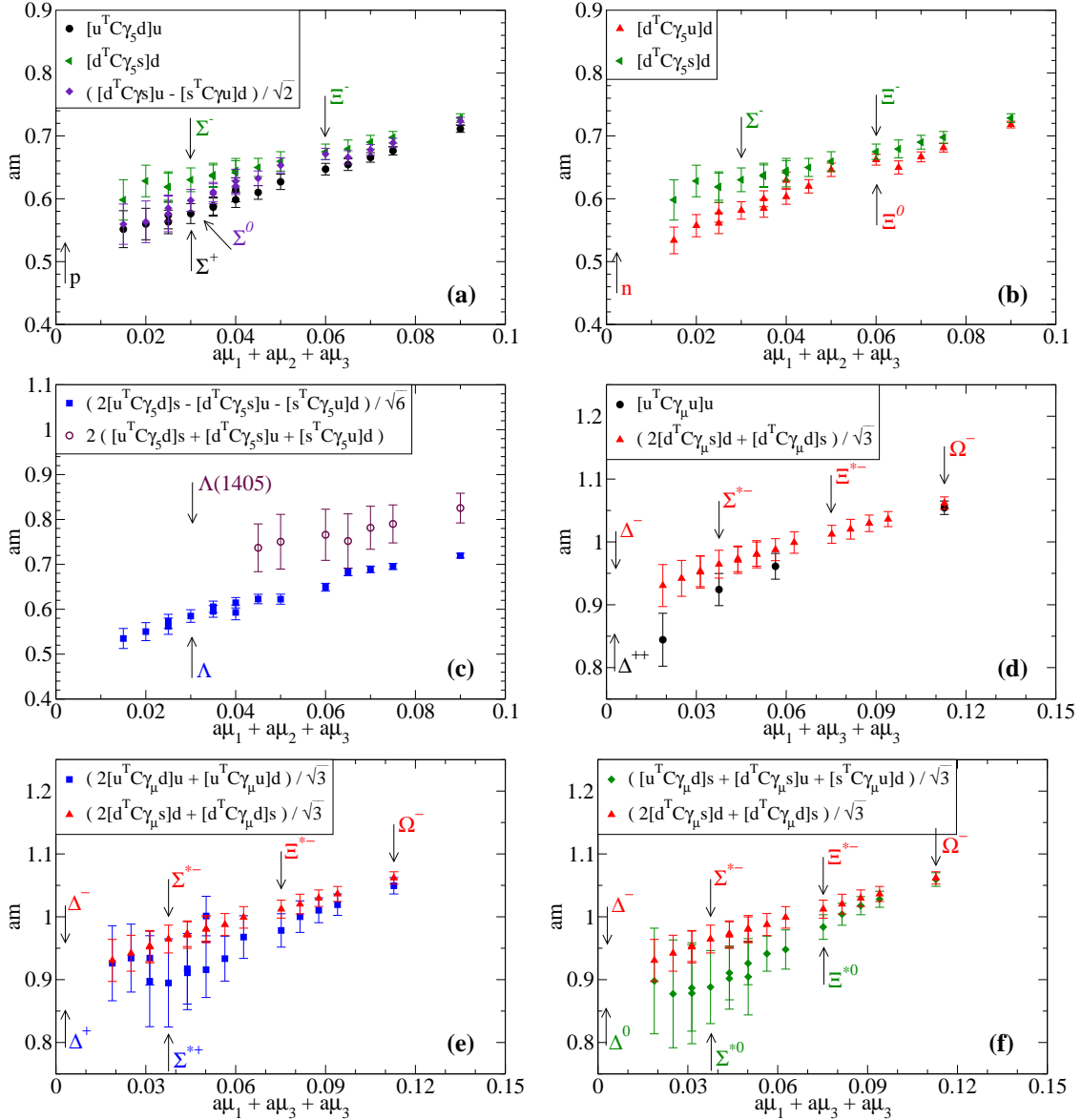

\scalebox{0.29}{\includegraphics*[0mm,11mm][257mm,19cm]{octet1b.eps}}
\scalebox{0.29}{\includegraphics*[0mm,11mm][257mm,19cm]{octet2b.eps}}
\scalebox{0.29}{\includegraphics*[0mm,11mm][257mm,19cm]{octet3b.eps}}
\scalebox{0.29}{\includegraphics*[0mm,11mm][257mm,19cm]{decuplet1b.eps}}
\scalebox{0.29}{\includegraphics*[0mm,11mm][257mm,19cm]{decuplet2b.eps}}
\scalebox{0.29}{\includegraphics*[0mm,11mm][257mm,19cm]{decuplet3b.eps}}
\caption{{\bf Panels (a), (b) and (c):} $(1/2)^+$ octet baryon masses, plus one
         $(1/2)^-$ channel to discuss the $\Lambda(1405)$, as functions of
         the sum of the twisted quark masses at $\beta=6.0$.
         {\bf Panels (d), (e) and (f):} $(3/2)^+$ decuplet baryon masses as
         functions of the sum of the twisted quark masses at $\beta=5.85$.
        }\label{baryonplots}
\end{figure}

The lack of flavour symmetry in tmLQCD (with $a\neq0$) means the physical
octet of spin-1/2 baryons has no exact mass degeneracies and
the decuplet of spin-3/2 baryons has just
two:\cite{Abdel-Rehim:2005gz,Walker-Loud:2005bt}
\begin{eqnarray}
m_{\Delta^{++}} &=& m_{\Delta^-}, \\
m_{\Delta^0} &=& m_{\Delta^+}.
\end{eqnarray}
For our numerical study of baryon masses,
the local source and sink operators are
given in the legends of each panel in Fig.~\ref{baryonplots}.
All combinations of quark masses from Table~\ref{latticedetails} were used
at each $\beta$ value for both spin-1/2 and spin-3/2 operators, though
Fig.~\ref{baryonplots} only displays data from
one $\beta$ as an example for each spin.

Any flavour splittings among these numerical results for baryons are
at the edge of statistical significance.
Clarification of these splittings will require
better operators and increased statistics.

\begin{figure}
\scalebox{0.63}{\includegraphics*[10mm,105mm][25cm,19cm]{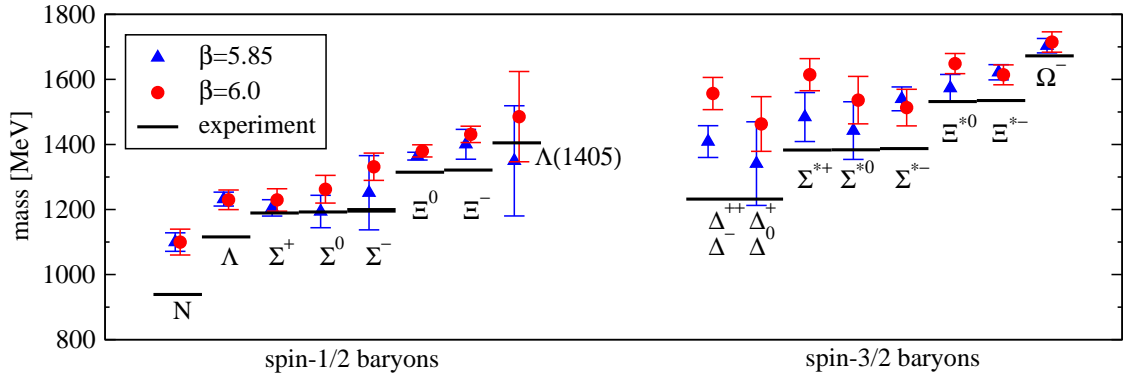}}
\caption{The quenched baryon spectrum compared to experiment.}\label{baryspec}
\end{figure}

Figure~\ref{baryspec} compares the baryon masses in physical units
to the spectrum observed in nature.  The tmLQCD results lie systematically
above experiment, which could be a quenching artifact and/or a consequence
of extracting the lattice spacing from $r_0$.  The two $\beta$ values produce
essentially equivalent results, suggesting that we are already in the scaling
region.

\section{Conclusions}

One particular method for incorporating the strange quark into tmLQCD
has been explored by directly computing the hadron mass spectrum.
This approach to 3-flavour tmLQCD produces a large splitting among kaon
masses at $a\gtrsim0.1$ fm.  The splitting vanishes linearly with $a^2$
as $a\to0$ as expected.
Some of the anticipated baryon splittings are potentially discernable, though
small.
Because the fermion determinant is complex, this method is
better suited for quenched simulations than for dynamical ones.
Other options for tmLQCD with a strange quark should also be investigated and
compared.

\section{Acknowledgements}

This work was supported in part by the Natural Sciences and Engineering
Research Council of Canada, the Canada Foundation for Innovation, the
Canada Research Chairs Program and the Government of Saskatchewan.

\providecommand{\href}[2]{#2}\begingroup\raggedright

\end{document}